%% file: Single_RR_fairness_empirical_study.tex
\documentclass[12pt]{article}
\usepackage[utf8]{inputenc}
\usepackage[british]{babel}
\usepackage{cmap}
\usepackage{lmodern}
\usepackage[T1]{fontenc}

\usepackage{amssymb, amsmath, amsthm}
\usepackage[a4paper,top=25mm,bottom=25mm,left=25mm,right=25mm]{geometry}
\usepackage{ragged2e}

\usepackage{authblk} % for headings
\usepackage{pifont}
\usepackage{graphicx}
\usepackage[dvipsnames,svgnames,table]{xcolor}
\usepackage[figuresright]{rotating}
\usepackage{xtab} % tackle the long tables
\usepackage{longtable} % tackle the long tables
\usepackage{multirow}
\usepackage{footnote}
\usepackage[stable]{footmisc}
\usepackage{chngpage} % allows for temporary adjustment of side margins
\usepackage{pdflscape} % landscape environment
\usepackage[nottoc,notlot,notlof]{tocbibind} % includes everything in ToC

\usepackage{pgfplots}
\pgfplotsset{every tick label/.append style={font=\footnotesize}}
\pgfplotsset{compat=1.18}
\usepackage{setspace}

\usepackage{array}
\newcolumntype{K}[1]{>{\centering\arraybackslash$}p{#1}<{$}}

\makesavenoteenv{tabular}
\usepackage{tabularx}
\usepackage{booktabs}
\usepackage{threeparttable}
\usepackage[referable]{threeparttablex} % footnotes in tabu
\newcolumntype{R}{>{\raggedleft\arraybackslash}X}
\newcolumntype{L}{>{\raggedright\arraybackslash}X}
\newcolumntype{C}{>{\centering\arraybackslash}X}
\newcolumntype{A}{>{\columncolor{gray!25}}C}
\newcolumntype{a}{>{\columncolor{gray!25}}c}

\newlength{\tablen}

\usepackage{dcolumn} % alignment to decimal points
\newcolumntype{.}{D{.}{.}{-1}}

\usepackage{tikz}
\usetikzlibrary{arrows, calc, matrix, patterns, positioning, trees}
\usepackage[semicolon]{natbib} % numbers,
\usepackage[hyphens]{xurl}
\usepackage{microtype}
\usepackage[justification=centering]{caption} % [justification=centering]

\usepackage[labelformat=simple]{subcaption}

\DeclareCaptionLabelFormat{parenthesis}{(#2)}
\makeatletter
\renewcommand\p@subfigure{\arabic{figure}.}
\makeatother

\DeclareCaptionLabelFormat{parenthesis}{(#2)}
\makeatletter
\renewcommand\p@subtable{\arabic{table}.}
\makeatother

\usepackage{enumitem}

\setlist[itemize]{leftmargin=2.5\parindent}
\setlist[enumerate]{leftmargin=2.5\parindent}

\usepackage{hyperref} % [hidelinks]
\hypersetup{
  colorlinks   = true,    		% Colours links instead of ugly boxes
  urlcolor     = blue,    		% Colour for external hyperlinks
  linkcolor    = blue,    		% Colour of internal links
  citecolor    = ForestGreen	% Colour of citations
}

\def\addlegendimage{\csname pgfplots@addlegendimage\endcsname}

\theoremstyle{plain}
\theoremstyle{definition}
\newtheorem{definition}{Definition}%[section]
\theoremstyle{remark}

\makeatletter
\let\@fnsymbol\@alph
\makeatother

\def\keywords{\vspace{.5em} % Add keywords
{\noindent \textit{Keywords}: }}

\def\AMS{\vspace{.5em} % Add keywords
{\noindent \textbf{\emph{MSC} class}: }}

\def\JEL{\vspace{.5em} % Add keywords
{\noindent \textbf{\emph{JEL} classification number}: }}

\title{On the fairness of single round-robin tournaments: Evidence from European football}
\author{
\href{https://sites.google.com/view/laszlocsato}{L\'aszl\'o Csat\'o}\thanks{~Corresponding author \newline
E-mail: \emph{laszlo.csato@sztaki.hun-ren.hu} \newline
Institute for Computer Science and Control (SZTAKI), Hungarian Research Network (HUN-REN), Laboratory on Engineering and Management Intelligence, Research Group of Operations Research and Decision Systems, Budapest, Hungary \newline % Hungary, 1111 Budapest, Kende street 13-17.
Corvinus University of Budapest (BCE), Institute of Operations and Decision Sciences, Department of Operations Research and Actuarial Sciences, Budapest, Hungary}
$\qquad \qquad$
Di\'ana Ivett F{\H u}r\'esz\thanks{~E-mail: \emph{furesz.diana@ktk.pte.hu} \newline
University of P\'ecs, Faculty of Business and Economics, P\'ecs, Hungary}
$\qquad \qquad$
Zsolt Havran\thanks{~E-mail: \emph{zsolt.havran@uni-corvinus.hu} \newline
Corvinus University of Budapest (BCE), Institute of Entrepreneurship and Innovation, Department of Business Economics, Budapest, Hungary}
}

\date{\today}

\def\Dedication{
{\noindent
``\emph{When accepting the presence of a home advantage in a single round robin context, where each team has a certain strength, the following is true: the particular set of teams played at home, and teams played away impacts the expected number of points or wins a team can collect. By that, the outcome of the tournament depends on which teams one plays home, and which teams one plays away.}''
\citep[p.~830]{WesselHurkensSpieksma2026}
}
\vspace{0.5cm}
\justify }

\begin{document}

\maketitle
\thispagestyle{empty}
\Dedication

\begin{abstract}
\noindent
Single round-robin tournaments are widely regarded as unfair because of their home-away imbalance: some competitors play more home games than the others. Therefore, most sports leagues are played in a double round-robin format. However, some European football leagues contain a single round-robin (sub)tournament. To test whether empirical data can validate their unfairness, we construct a unique database and develop a strength of schedule metric based on Elo ratings with favourable theoretical properties. In contrast to the robust impact of relative team strength, the strength of schedule is found to have no significant effect on the number of points scored under the assumption of a league-specific home advantage. Consequently, the theoretical unfairness is not a serious issue in practice, or the standard approach of modelling home advantage in Elo ratings might need to be reconsidered.
\end{abstract}


\AMS{62F07, 62P20, 90B90}
% Ranking and selection
% Applications of statistics to economics
% Case-oriented studies in operations research

\JEL{C44, C53, D71, Z20}
% Operations Research + Statistical Decision Theory 
% Forecasting and Prediction Methods + Simulation Methods 
% Social Choice, Clubs, Committees, Associations
% Sports Economics, General

\clearpage

\section*{Motivation} \label{Sec0}

The Hungarian first football division is organised in a triple round-robin format since the 2015/16 season: any given pair of teams $i$ and $j$ plays three times against each other such that either team $i$ or team $j$ plays twice at home. Hence, all the 12 teams play 33 matches, but six teams play 17 home matches, while the remaining six teams play only 16 home matches.

In the 2017/18 season, Balmaz\'ujv\'aros and Di\'osgy{\H o}r contested to avoid relegation. Both clubs finished with 36 points; Balmaz\'ujv\'aros played fewer home games and had a superior goal difference -- but was relegated as the primary tie-breaking rule is the number of wins.
In the 2025/26 season, after seven consecutive triumphs for Ferencv\'aros, Gy{\H o}r became the champion ahead of Ferencv\'aros with the minimal advantage of one point; the tie-breaking rules favoured Ferencv\'aros. However, Gy{\H o}r played twice at home against Ferencv\'aros.

One can argue that the structure of the league affected these outcomes, and might have been responsible for the relegation of Balmaz\'ujv\'aros and the disappointment of Ferencv\'aros. The current paper aims to empirically explore the impact of this inherent bias in single round-robin tournaments, caused by the asymmetric allocation of home advantages.

\section{Introduction} \label{Sec1}

Every professional sports league produces a ranking at the end of the season. Thus, one of the most important responsibilities of decision-makers is guaranteeing that the rules of the competition do not affect the ranking, which solely depends on performance on the field. In particular, if the ranking aggregates the results of matches, all teams should meet opponents of comparable strength. This explains the popularity of the round-robin format, where each team plays the same number of games against any other team.

Obviously, asymmetry in the games can be balanced only if every pair of teams plays an even number of games against each other. Asymmetry arises in many sports. For instance, playing with the white pieces is considered beneficial in chess \citep{Henery1992, Fecker2024}. Analogously, if the match is played at the venue of one of the teams, the well-documented home advantage \citep{PollardPollard2005} implies asymmetry. Unsurprisingly, most European football leagues use a double round-robin format \citep{GoossensSpieksma2012b, LasekGagolewski2018}, since home advantage plays an important role in determining the result of a game \citep{Pollard1986, Pollard2006, Pollard2008}.

However, some leagues contain single round-robin (sub)tournaments as mentioned above.
A recent theoretical paper, \citet{WesselHurkensSpieksma2026}, argues that each player should face its opponents when ranked by strength in an alternating fashion with respect to home advantage to prevent distorting the outcome and to guarantee equal treatment. First in the literature, we check whether this claim of distortion is supported by empirical data.

In particular, our paper aims to study the fairness of single round-robin competitions by examining the following research question: \emph{Does the number of points scored depend on the allocation of home and away games?}
In order to statistically test this issue, a unique database is constructed from European football leagues (Section~\ref{Sec3}). We define average winning probability based on Elo ratings and develop a strength of schedule metric (Section~\ref{Sec4}). Our measure uses a constant home advantage parameter (which is the standard used by Elo ratings in football), equals zero in the absence of home advantage, and takes into account the relative strength of home and away opponents. The latter is a key feature because, for example, a strong team is severely disadvantaged if it plays against the strongest half of its opponents away, and against the weakest half of its opponents at home \citep{WesselHurkensSpieksma2026}.
The statistical analysis applies the correlated random-effects method based on the Mundlak specification.

The results in Section~\ref{Sec5} consistently show that, under the assumption of home advantage enjoyed equally by all teams across the league, strength of schedule does not have any statistically significant impact on the number of points. Thus, the allocation of home games -- including the imbalance in the number of home and away games -- seems not to make these tournaments unfair in practice, even though home advantage can be clearly observed.

%The study has the following structure. Section~\ref{Sec2} provides a concise overview of the previous literature. The data are discussed in Section~\ref{Sec3} and the methodology is described in Section~\ref{Sec4}. Section~\ref{Sec5} presents the results and Section~\ref{Sec6} discusses them.

\section{Related literature} \label{Sec2}

As far as we know, no previous study has examined the fairness of single round-robin tournaments empirically, in contrast to the impact of the schedule in double round-robin tournaments and competitions played on a neutral field.
If team A plays against team B, and subsequently against team C, the performance of team A against team C may depend on team B; hence, team C is said to receive a carryover effect from team B. \citet{GoossensSpieksma2012a} develop a method to measure whether the carryover effect influences the outcome of football matches, and apply it to the highest Belgian division. No supporting evidence is found, implying that the carryover effect is not a major issue in sports scheduling from the perspective of league organisers, even though it plays a central role in the scheduling literature \citep{LambrechtsFickerGoossensSpieksma2018, GoossensYiVanBulck2020, DevriesereCsatoGoossens2025}.
\citet{KrumerLechner2017} study the fairness of sequential round-robin tournaments with three and four contestants. According to their results based on Olympic wrestling (three athletes per group), as well as FIFA World Cups and UEFA European Championships (four teams per group), competing in the first and third matches provides a significant advantage.

\citet{Tuu-SzaboKoczy2026} propose a Monte Carlo simulation framework to quantify procedural fairness in league competitions by isolating biases caused by fixture schedules from team strength and ranking rules. They show that unadjusted triple round-robin tournaments, such as the Hungarian first football division, create substantially larger and less evenly distributed fairness distortions, while balancing mechanisms, like those used in Switzerland, can substantially reduce these effects.
%propose a simulation-based approach to quantify schedule-induced distortions in final league standings. The model is applied to two European football leagues with a triple round-robin format, the Hungarian National League (NBI) and the Swiss Super League.

\citet{WesselHurkensSpieksma2026} introduce a novel fairness measure for a single round-robin tournament where the contestants are ranked by their strength. In order to guarantee equal treatment, each contestant should face its opponents when ranked by strength in an alternating pattern with respect to home advantage. The existence of ranking-fair schedules is proved if the number of contestants is a multiple of four, and an algorithm is given to construct them.

A potential problem of fairness, analogous to single round-robin competitions, arises in Swiss-system tournaments with an odd number of rounds. \citet{CsatoKrumer2026} compare the number of points scored in Swiss-system chess tournaments by two groups of players, who played one game more/less with the white pieces. Players with an extra white game score significantly more points; the advantage exceeds the value of a draw in the prestigious FIDE Grand Swiss tournaments played in 11 rounds.

Unbalanced schedules have received serious attention in sports economics since they can distort standard measures of competitive balance.
\citet{Lenten2008} suggests a log-probability rule to derive adjusted win ratios correcting for the bias.
\citet{Lenten2011} uses a modified model and applies it to the Australian Football League (AFL) to create an adjusted league ranking.
\citet{Lenten2015} extends this model to the conference and divisional design of the National Football League (NFL).
\citet{LeeKimKim2019} show that the winning percentage is a biased estimator of the true win probability if schedules are unbalanced, and present a mathematical statistical approach to compute an unbiased estimator of within-season competitive balance.
However, none of these studies test whether the implied bias can be empirically detected in the league rankings, which will be done in the following for single round-robin (sub)tournaments in European football.

Finally, we know only one paper about scheduling real-world single round-robin tournaments. \citet{Rasmussen2008} proposes a solution method based on a logic-based Benders decomposition to find a schedule for the triple round-robin tournament played in Denmark. The given algorithm works for practical applications, but struggles to prove optimality. It was adopted by the Danish Football Association to schedule the 2006/07 season.

\section{Data} \label{Sec3}

While the top European football leagues have been organised in a double round-robin format in recent decades, there exist some exceptions among second-tier leagues that can be considered to explore the fairness of single round-robin competitions. In the absence of any comprehensive database on the format of these leagues, we have managed to find the following candidates to construct a database.

Three countries have recently used a triple round-robin format with 12 teams such that the schedule of the whole season was determined before the first round:
\begin{itemize}
\item
Denmark: between 1995/96 and 2015/16 (21 seasons);
\item
Hungary: between 2015/16 and 2025/26 (11 seasons);
\item
Scotland (before the split): between 2000/01 and 2025/26, except for the unfinished 2019/20 season (25 seasons).
\end{itemize}
Three West Balkan countries have recently used a triple round-robin with 12 teams such that the schedule of the last 11 rounds was determined before the 23rd round, based on the positions after the first 22 rounds:
\begin{itemize} 
\item
Croatia: 1999/2000, between 2006/07 and 2008/09, 2012/13 (5 seasons);
% \url{https://en.wikipedia.org/wiki/1999\%E2\%80\%932000_Croatian_First_Football_League#Results}
\item
Serbia: between 2007/08 and 2008/09 (2 seasons);
% \url{https://en.wikipedia.org/wiki/2008\%E2\%80\%9309_Serbian_SuperLiga#Third_round}
\item
Slovenia: between 1998/99 and 2002/03 (5 seasons).
% \url{https://en.wikipedia.org/wiki/2002\%E2\%80\%9303_Slovenian_PrvaLiga})
\end{itemize}
Finally, two nations have used a league system with 16 teams where a subset of teams play an additional single round-robin tournament after a double round-robin:
\begin{itemize}
\item
Romania: the last ten teams since 2020/21 (6 seasons);
\item
Serbia: the first eight teams (championship part) and the last eight teams (relegation part) since 2015/16, except for the unfinished 2019/20 and 2020/21 seasons (9 seasons).
\end{itemize}
Based on the list above, we have chosen the countries where a single round-robin tournament is organised at least over 10 years: Denmark (1995/96--2015/16), Hungary (since 20215/16), Scotland (since 2000/01), and Serbia (since 2015/16).

In order to estimate the potential bias caused by the allocation of home games in a single round-robin tournament, it is crucial to have a reliable measure of team strengths. For this purpose, we use Football Club Elo Ratings (\url{http://clubelo.com/}), a well-established measure in the academic literature \citep{BoskerGurtler2024, Csato2022b, Csato2024c, CsatoPetroczy2025a, CsatoPetroczy2026, YildirimBilman2025a, YildirimBilman2025b}.
Even though the triple round-robin format has been used recently in further leagues, such as the Faroe Islands, Finland, Israel, Kosovo, and Slovakia, they have not been included in the sample because even the top divisions of these countries are missing from the database of Football Club Elo Ratings.

The basic idea behind the Elo method is straightforward: the ratings of the two opponents are updated based on the outcome of their match; the same result against a stronger (weaker) opponent leads to a higher (smaller) gain, and the impact of a game decreases after new games are played \citep{CsatoKissSzadoczki2025}.
In contrast to the official FIFA World Ranking, which uses the Elo approach since 2018 \citep{FIFA2018c}, Football Club Elo Ratings take both home advantage and goal difference into account as recommended by \citet{SzczecinskiRoatis2022}. Home advantage is a league-specific, dynamic parameter; its value increases or decreases together with the Elo rating of the home team. Hence, Football Club Elo Ratings are able to capture trends in home advantage. Unsurprisingly, this parameter in the top European leagues reached its minimum in the seasons affected by the Covid-19 pandemic when the matches were played behind closed doors.

Elo rating is a strictly performance-based measure that does not reflect the effect of injuries, managerial changes, or player transfers. Hence, Elo ratings are usually not competitive with betting odds in forecasting accuracy \citep{Wunderlich2025}.
Furthermore, the database of Football Club Elo Ratings does not contain the second football divisions of these countries. Thus, the ratings of the promoted teams are equal at the beginning of a season and are an unreliable measure of strength.

To conclude, it would be suboptimal using pre-season Elo ratings rather than Elo ratings after the standard double round-robin part of the league is played. This does not cause any problem in Hungary, Scotland, and Serbia except for some marginal corrections detailed in the Appendix.
However, the Danish league started with a single round-robin of 11 rounds, followed by a double round-robin of 22 rounds \citep{Rasmussen2008}, which excludes using the Danish data.

Therefore, we construct club-level panel data from three European football leagues (Hungary, Scotland, Serbia). Although observations for the Scottish league are available from the 2000/01 season, analogous data for the Hungarian and Serbian leagues do not exist before the 2015/16 season. To ensure temporal comparability, the primary sample is restricted to the common observation period from 2015/16 to 2025/26.
The exclusion of Scottish data prior to the 2015/16 season is particularly important given the cross-country focus, as retaining the earlier observations would make country membership partially confounded with the observation period. Restricting the analysis to a common time window ensures a consistent basis for estimating country effects and country-specific interaction terms.
Nonetheless, we consider all observations from Scotland in the secondary sample.

\begin{table}[t!]
  \centering
  \caption{Sample coverage}
  \label{Table1}

\begin{subtable}{\textwidth}
\centering
  \caption{Primary sample (2025/16--2025/26)}
  \label{Table1a}
\begin{threeparttable}
    \rowcolors{1}{}{gray!20}
    \begin{tabularx}{0.9\textwidth}{lCcCC} \toprule
    Country & Seasons & Teams per season & Observations & Clubs \\ \bottomrule
    Hungary & 11    & 12    & 132   & 23 \\
    Scotland & 10    & 12    & 120   & 17 \\
    Serbia & 9     & $2 \times 8$ & 144   & 31 \\ \toprule
    Total & 30    & ---   & 396   & 71 \\ \bottomrule
    \end{tabularx}
\begin{tablenotes} \footnotesize
\item
\emph{Notes}:
Hungary: All seasons from 2015/16 to 2025/26. \\
Scotland: All seasons from 2015/16 to 2025/26, except for the unfinished 2019/20. \\
Serbia: All seasons from 2015/16 to 2025/26, except for the unfinished 2019/20 and 2020/21. 
\end{tablenotes}
\end{threeparttable}
\end{subtable}

\vspace{0.5cm}
\begin{subtable}{\textwidth}
\centering
  \caption{Secondary sample (Scotland)}
  \label{Table1b}
\begin{threeparttable}
    %\rowcolors{1}{}{gray!20}
    \begin{tabularx}{0.9\textwidth}{lCcCC} \toprule
    Country & Seasons & Teams per season & Observations & Clubs \\ \midrule
    Scotland & 25    & 12    & 300 & 19 \\ \bottomrule
    \end{tabularx}
\begin{tablenotes} \footnotesize
\item
\emph{Note}:
All seasons from 2000/01 to 2025/26, except for the unfinished 2019/20.
\end{tablenotes}
\end{threeparttable}
\end{subtable}

\end{table}

The two samples are summarised in Table~\ref{Table1}.
Both constitute an unbalanced panel: individual clubs may enter or leave the sample over time due to promotion and relegation. The club is the cross-sectional unit and season is the time dimension.

We use Football Club Elo Ratings as of the first day of Round 23 in the Hungarian and Scottish leagues, and as of the first day of Round 31 in the Serbian league. The exact dates are provided in Table~\ref{Table_A1} in the Appendix.

\section{Methodology} \label{Sec4}

The background of the estimation is described in two parts.
Section~\ref{Sec41} presents how we control for the strength of opponents and the strength of schedule, while Section~\ref{Sec42} discusses the selection of the statistical approach.

\subsection{Variable and model construction} \label{Sec41}

Let $R_i$ and $R_j$ denote the Elo ratings of teams $i$ and $j$, respectively.
According to the methodology of Football Club Elo Ratings, team $i$ wins against team $j$ at home with a probability of
\begin{equation} \label{eq_Elo}
p_{ij}(\Delta) = \frac{1}{1 + 10^{\left( R_j - R_i - \Delta \right)/400}},
\end{equation}
where $\Delta$ is the home advantage parameter (the Elo rating of the home team is increased by $\Delta$).
Since the Elo method does not consider draws in the calculation of winning probabilities, the probability that team $j$ wins against team $i$ away equals
\[
1- p_{ij}(\Delta) = 1 - \frac{1}{1 + 10^{\left( R_j - R_i - \Delta \right)/400}} = \frac{1}{1 + 10^{\left( R_i - R_j + \Delta \right)/400}}.
\]

The Elo ratings are updated after the match between teams $i$ and $j$ as follows:
\[
R'_i =  R_i + k \left[ Q_{ij} - p_{ij}(\Delta) \right] \qquad \text{and} \qquad R'_j =  R_j + k \left[ \left( 1 - Q_{ij} \right) - \left( 1 - p_{ij}(\Delta) \right) \right],
\]
where $Q_{ij}$ is the outcome of the match; $Q_{ij} = 1$ if team $i$ wins, $Q_{ij} = 0$ if team $j$ wins, and $Q_{ij} = 0.5$ if they play a draw.
Consequently, $R_i + R_j = R'_i + R'_j$. Football Club Elo Ratings assume $k=20$.

Let $N$ be the set of teams and $\lvert N \rvert = n$ be the number of teams in a tournament.
The logic behind the Elo method suggests the following definition.

\begin{definition} \label{Def1}
The \emph{average winning probability} of team $i$ in a double round-robin tournament is
\begin{equation} \label{eq_AWP}
\mathit{AWP}_i(\Delta) = \sum_{j \in N, j \neq i} \frac{p_{ij}(\Delta) + p_{ji}(\Delta)}{2(n-1)}.
\end{equation}
\end{definition}
\noindent
$\mathit{AWP}_i(\Delta)$ is the mean of winning probabilities from equation~\eqref{eq_Elo}, taken over the set of matches played against all other teams, once at home and once away.

$\mathit{AWP}_i(\Delta) = 0.5$ for each team $i \in N$ if they are of equal strength.
In the case of competitive imbalance, $\mathit{AWP}_i(\Delta)$ is influenced by the value of home advantage $\Delta$ due to the non-linearity of $p_{ij}$ as a function of $R_j - R_i$. However, the dependence is rather marginal because the reduction in the probability of winning away is balanced by the increase in the probability of winning at home to a great extent.

Denote by $O_i^h$ ($O_i^a$) the opponents of team $i$ against whom it plays at home (away) in a tournament.

\begin{definition} \label{Def2}
The \emph{strength of schedule} of team $i$ in a tournament is
\begin{equation} \label{eq_SoS}
\mathit{SoS}_i(\Delta) = 100 \times \left( \frac{\sum_{j \in O_i^h} p_{ij}(\Delta) + \sum_{j \in O_i^a} p_{ji}(\Delta)}{\left| O_i^h \right| + \left| O_i^a \right|} - \mathit{AWP}_i(\Delta) \right).
\end{equation}
\end{definition}
\noindent
This metric, inspired by equation~\eqref{eq_AWP}, is the difference between the average winning probabilities in the tournament considered and a (fair) double round-robin tournament, respectively, multiplied by 100.
In a single round-robin tournament without home advantage ($\Delta = 0$), as well as in a double round-robin, $\mathit{SoS}_i(\Delta) = 0$ for each team $i \in N$.

\begin{table}[t!]
  \centering
  \caption{League-specific home advantage parameters}
  \label{Table2}

\begin{threeparttable}
    %\rowcolors{1}{}{gray!20}
    \begin{tabularx}{0.7\textwidth}{lCCC} \toprule
    Country & End of sample & Minimum & Maximum \\ \midrule
    \multirow{2}[0]{*}{Hungary} & 46.6  & 42.1  & 156.5 \\
          & (2026-06-30) & (2026-03-21) & (1980-03-01) \\
    \multirow{2}[0]{*}{Scotland} & 66.5  & 21.9  & 113.5 \\
          & (2026-06-30) & (2017-12-09) & (1953-10-31) \\
    \multirow{2}[0]{*}{Serbia} & 47.9  & 45.4  & 101.5 \\
          & (2026-06-30) & (2025-08-18) & (2007-05-30) \\ \bottomrule
    \end{tabularx}
\begin{tablenotes} \footnotesize
\item
\emph{Source}: Football Club Elo Ratings (\url{http://clubelo.com/}). 
\end{tablenotes}
\end{threeparttable}
\end{table}

Since we use ex-ante Elo ratings, both the average winning probability $\mathit{AWP}$ and the strength of schedule $\mathit{SoS}$ are independent of the result of any match played in the single round-robin (sub)tournament of the given season.
On the other hand, both $\mathit{AWP}$ and $\mathit{SoS}$ depend on the home advantage parameter $\Delta$.
Some league-specific home advantage values of Football Club Elo Ratings are shown in Table~\ref{Table2}. $\Delta = 50$ is chosen as a benchmark, while $\Delta = 25$ and $\Delta = 75$ are used for sensitivity analyses.

Our empirical analysis examines whether schedule characteristics are associated with sporting performance in single round-robin competitions. The unit of observation is the club-season, resulting in an unbalanced panel of clubs from Hungary, Serbia, and Scotland over the 2015/16--2025/26 period. The dependent variable is the number of points per match rather than accumulated points. This normalisation is important because the number of matches underlying the scheduling measures differs across the countries.

The baseline relationship is expressed as follows:
\[
S_{it} = \alpha + \beta_1 \mathit{AWP}_{it} + \beta_2 \mathit{SoS}_{it} + \gamma_1 \mathit{HUN}_i + \gamma_2 \mathit{SRB}_i + \varepsilon_{it},
\]
where $i$ denotes the clubs and $t$ denotes the seasons.
The dependent variable $S_{it}$ is the number of points collected per match. $\mathit{AWP}_{it}$ is the average winning probability, $\mathit{SoS}_{it}$ represents the strength of schedule. Scotland serves as the reference country; the dummy variables $\mathit{HUN}_i$ and $\mathit{SRB}_i$ identify Hungarian and Serbian clubs, respectively.
%Note that $\mathit{AWP}_{it}$ and $\mathit{SoS}_{it}$ depend on the home advantage parameter $\Delta$.

\begin{table}[t!]
  \centering
  \caption{Summary statistics of the main variables}
  \label{Table3}

\begin{subtable}{\textwidth}
\centering
  \caption{Primary sample (2015/16--2025/26)}
  \label{Table3a}
    \rowcolors{1}{}{gray!20}
    \begin{tabularx}{0.9\textwidth}{lcCC} \toprule
    Measure & $S$ (points per match) & $\mathit{AWP} (\Delta = 50)$ & $\mathit{SoS} (\Delta = 50)$ \\ \bottomrule
    Mean  & 1.38  & 0.50  & \textcolor{gray!20}{$+$}0.00 \\
    St.~dev. & 0.56  & 0.13  & \textcolor{white}{$+$}0.83 \\
    Minimum & 0     & 0.25  & $-$2.15 \\
    Median & 1.29  & 0.47  & \textcolor{white}{$+$}0.04 \\
    Maximum & 3     & 0.90  & \textcolor{gray!20}{$+$}2.15 \\ \toprule
    \end{tabularx}
\end{subtable}

\vspace{0.5cm}
\begin{subtable}{\textwidth}
\centering
  \caption{Secondary sample (Scotland)}
  \label{Table3b}
    \rowcolors{1}{}{gray!20}
    \begin{tabularx}{0.9\textwidth}{lcCC} \toprule
    Measure & $S$ (points per match) & $\mathit{AWP} (\Delta = 50)$ & $\mathit{SoS} (\Delta = 50)$ \\ \bottomrule
    Mean  & 1.38  & 0.50  & \textcolor{gray!20}{$+$}0.00 \\
    St. dev. & 0.58  & 0.15  & \textcolor{white}{$+$}0.63 \\
    Minimum & 0.27  & 0.22  & $-$1.27 \\
    Median & 1.27  & 0.46  & \textcolor{white}{$+$}0.01 \\
    Maximum & 2.82  & 0.90  & \textcolor{gray!20}{$+$}1.11 \\ \toprule
    \end{tabularx}
\end{subtable}

\end{table}

Table~\ref{Table3} reports the main descriptive statistics of the two variables constructed, together with the number of points per match.
The mean average winning probability is almost exactly 0.5, and the mean strength of schedule is essentially zero. Average winning probability varies roughly between 25\% and 90\%. Accounting for the allocation of home matches biases this probability by no more than 2.5 percentage points in the baseline model with home advantage $\Delta = 50$.

\subsection{Selection of the estimation method} \label{Sec42}

\begin{table}[t!]
\centering
  \caption{Model specification tests, primary sample (2015/16--2025/26)}
  \label{Table4}
\begin{threeparttable}
    \rowcolors{1}{}{gray!20}
    \begin{tabularx}{0.9\textwidth}{lCC} \toprule
    Test  & Statistic & $p$-value \\ \bottomrule
    FE vs.\ pooled OLS F-test & $F(70, 323) = 1.276$ & 0.084 \\
    RE vs.\ pooled OLS BP-LM test & $\chi^2(1) = 0.101$ & 0.751 \\
    Hausman test, FE vs.\ RE & $\chi^2(2) = 13.930$ & $<$ 0.001 \\ \toprule
    \end{tabularx}
\begin{tablenotes} \footnotesize
\item
\emph{Notes}:
Baseline specification with home advantage parameter $\Delta = 50$.
\item
BP-LM = Breusch–Pagan Lagrange multiplier; FE = fixed effects; RE = random effects.
\end{tablenotes}
\end{threeparttable}
\end{table}

Since the dataset has a panel structure, we initially considered pooled ordinary least squares (OLS), fixed-effects (FE), and random-effects (RE) specifications. As reported in Table~\ref{Table4}, the FE versus pooled OLS F-test does not reject the pooled specification. Similarly, the Breusch--Pagan Lagrange Multiplier test provides no evidence in favour of RE over pooled OLS. The Hausman test rejects the conventional RE orthogonality assumption. Taken together, these tests do not justify selecting the conventional RE estimator and do not reject pooled OLS in favour of a model with unit-specific effects at the conventional 5\% significance level.

Nevertheless, distinguishing within-club from between-club variation is important because schedule characteristics may vary both across seasons for the same club and systematically across clubs. The pooled OLS specification was therefore augmented with club-specific temporal means of the time-varying regressors following the Mundlak approach \citep{Mundlak1978, Wooldridge2002}. This implementation is consistent with the formulation by \citet{Wooldridge2025} of Mundlak regressions, using pooled OLS with unit-specific time averages as additional regressors.

Thus, the estimated specification is
\[
S_{it} = \beta_0 + \beta_1 \mathit{AWP}_{it} + \beta_2 \mathit{SoS}_{it} + \gamma_1 \mathit{HUN}_i + \gamma_2 \mathit{SRB}_i + \theta_1 \overline{\mathit{AWP}}_i + \theta_2 \overline{\mathit{SoS}}_i +  \varepsilon_{it},
\]
where the terms $\overline{\mathit{AWP}}_i$ and $\overline{\mathit{SoS}}_i$ are club-specific temporal means of the corresponding time-varying regressors and constitute the Mundlak correction.
%Including club-specific temporal means allows the coefficients on the contemporaneous time-varying regressors to be interpreted as within-club associations conditional on the corresponding club means.
Under this parameterisation, $\theta_1$ and $\theta_2$ capture the difference between the between-club and within-club slopes, such that the implied between-club slope for a variable is given by $\beta + \theta$.

Repeated observations from the same club may exhibit heteroskedasticity and within-club dependence; hence, statistical inference is based on HC1 robust standard errors clustered at the club level. Club-specific temporal means are jointly significant in the baseline and sensitivity specifications (see Section~\ref{Sec5}), supporting their inclusion in the final models.

%The model with $\mathit{AWP}(\Delta=50)$ and $\mathit{SoS}(\Delta=50)$ is our baseline specification.
The estimation is conducted in R using the \texttt{plm} (Linear Models for Panel Data) package (\url{https://cran.r-project.org/web/packages/plm/index.html}).

\section{Results} \label{Sec5}

Section~\ref{Sec51} offers some descriptive statistics for the three leagues in order to help understanding the main characteristics of the data. The results of the baseline statistical model are presented in Section~\ref{Sec52}, and sensitivity analyses are given in Section~\ref{Sec53}.

\subsection{Preliminary data analysis} \label{Sec51}

\begin{table}[t!]
  \centering
  \caption{Match outcomes in the single round-robin (sub)tournaments}
  \label{Table5}

\begin{subtable}{\textwidth}
\centering
  \caption{Hungary, 2015/16--2025/26}
  \label{Table4a}
    \rowcolors{1}{}{gray!20}
    \begin{tabularx}{0.8\textwidth}{lCCCc} \toprule
          & Home win & Draw & Away win & Home-away gap \\ \bottomrule    
    Average & 44.21\% & 27.55\% & 28.24\% & 15.98\% \\
    Minimum & 36.36\% & 21.21\% & 21.21\% & \textcolor{white}{0}4.55\% \\
    Maximum & 57.58\% & 37.88\% & 34.85\% & 36.36\% \\ \toprule
    \end{tabularx}
\end{subtable}

\vspace{0.5cm}
\begin{subtable}{\textwidth}
\centering
  \caption{Scotland, 2015/16--2025/26 (2019/20 excluded)}
  \label{Table4b}
    \rowcolors{1}{}{gray!20}
    \begin{tabularx}{0.8\textwidth}{lCCCc} \toprule
          & Home win & Draw & Away win & Home-away gap \\ \bottomrule
    Average & 45.61\% & 22.88\% & 31.52\% & 14.09\% \\
    Minimum & 40.91\% & 12.12\% & 21.21\% & 4.55\% \\
    Maximum & 53.03\% & 31.82\% & 36.36\% & 31.82\% \\ \toprule
    \end{tabularx}
\end{subtable}

\vspace{0.5cm}
\begin{subtable}{\textwidth}
\centering
  \caption{Serbia, 2015/16--2025/26 (2019/20 and 2020/21 excluded)}
  \label{Table4c}
\begin{threeparttable}
    \rowcolors{1}{}{gray!20}
    \begin{tabularx}{0.8\textwidth}{lCCCc} \toprule
          & Home win & Draw & Away win & Home-away gap \\ \bottomrule
    Average & 49.21\% & 23.41\% & 27.38\% & 21.83\% \\
    Minimum & 32.14\% & \textcolor{white}{0}3.57\% & 14.29\% & \textcolor{white}{0}3.57\% \\
    Maximum & 75.00\% & 46.43\% & 35.71\% & 53.57\% \\ \toprule
    \end{tabularx}
\begin{tablenotes} \footnotesize
\item
\emph{Note}:
The championship part, contested by the top eight teams after 30 rounds, and the relegation part, contested by the bottom eight teams after 30 rounds, are treated separately.
\end{tablenotes}
\end{threeparttable}
\end{subtable}

\vspace{0.5cm}
\begin{subtable}{\textwidth}
\centering
  \caption{Scotland, 2000/01--2025/26 (2019/20 excluded)}
  \label{Table4d}
    \rowcolors{1}{}{gray!20}
    \begin{tabularx}{0.8\textwidth}{lCCCc} \toprule
          & Home win & Draw & Away win & Home-away gap \\ \bottomrule
    Average & 45.52\% & 24.36\% & 30.12\% & \textcolor{gray!20}{$-$}15.39\% \\
    Minimum & 37.88\% & 12.12\% & 19.70\% & \textcolor{white}{0}$-$1.52\% \\
    Maximum & 53.03\% & 40.91\% & 40.91\% & \textcolor{gray!20}{$-$}31.82\% \\ \toprule
    \end{tabularx}
\end{subtable}
\end{table}

Table~\ref{Table5} summarises the descriptive statistics of trichotomous match outcomes for the three leagues. As expected, home teams consistently have a higher probability of winning than away teams; their advantage is around 15 percentage points in Hungary and Scotland, but exceeds 20 in Serbia. The reason is that the home-away patterns are fixed at the beginning of the season in the former two countries, while the first half of the teams after 30 rounds (ranked 1--4 in the championship part and 9--12 in the relegation part) are guaranteed to play more (four) home games in the Serbian league. Consequently, more home games are positively correlated with team strength in Serbia.
The range of match outcomes across the seasons is also higher in Serbia, explained by the higher randomness of single round-robin tournaments with eight teams compared to tournaments with 12 teams.

\begin{table}
\centering
  \caption{Share of points collected by the teams with more home games}
  \label{Table6}
\begin{threeparttable}
    \rowcolors{1}{}{gray!20}
    \begin{tabularx}{0.6\textwidth}{lCCC} \toprule
          & Hungary & Scotland & Serbia \\ \bottomrule
    Average & 50.11\% & 50.90\% & 56.91\% \\
    Minimum & 41.67\% & 42.93\% & 33.77\% \\
    Maximum & 58.70\% & 61.75\% & 71.43\% \\ \toprule
    \end{tabularx}
\begin{tablenotes} \footnotesize
\item
\emph{Notes}:
Hungary: All seasons from 2015/16 to 2025/26.
Scotland: All seasons from 2000/01 to 2025/26, except for the unfinished 2019/20.
Serbia: All seasons from 2015/16 to 2025/26, except for the unfinished 2019/20 and 2020/21. 
\end{tablenotes}
\end{threeparttable}
\end{table}

The share of teams playing more home games from the total number of points collected is presented in Table~\ref{Table6}. In Hungary and Scotland, this varies roughly between 40\% and 60\%, with an average of almost exactly 50\%. In contrast, the mean robustly exceeds 50\% in Serbia due to the scheduling policy that favours teams performing better in the first 30 rounds. The variance is also higher in Serbia, again because of the smaller number of teams in the single round-robin (sub)tournaments.

\input{Figure1_AWP_points_per_match}

Figure~\ref{Fig1} shows the relationship between our measure of relative team strength, the average winning probability $\mathit{AWP}$, and the dependent variable, the number of points per match $S$, in the estimated models, for the primary sample. A clear positive but, naturally, imperfect correlation can be seen. Extreme values in $S$ are more frequent in the Serbian league as its relevant (sub)tournament consists of seven rounds instead of 11.

\subsection{The baseline Mundlak-augmented pooled OLS estimation} \label{Sec52}

\begin{table}[t!]
\centering
  \caption{Mundlak-augmented pooled OLS estimates, primary sample (2015/16--2025/26)}
  \label{Table7}
\begin{threeparttable}
    \rowcolors{1}{gray!20}{}
    \begin{tabularx}{\textwidth}{lCCC} \toprule \hiderowcolors
    Model & Baseline & Sensitivity & Sensitivity \\
    Home advantage $\Delta$ & 50    & 25    & 75 \\ \midrule
    \multirow{2}[0]{*}{Constant} & $-$0.082 & $-$0.069 & $-$0.103 \\
          & (0.073) & (0.072) & (0.074) \\
    \multirow{2}[0]{*}{$\mathit{AWP}$} & 1.597*** & 1.575*** & 1.628*** \\
          & (0.318) & (0.315) & (0.323) \\
    \multirow{2}[0]{*}{$\mathit{SoS}$} & 0.006 & 0.014 & 0.004 \\
          & (0.034) & (0.068) & (0.023) \\
    \multirow{2}[0]{*}{Hungary} & $-$0.023 & $-$0.023 & $-$0.023 \\
          & (0.045) & (0.045) & (0.045) \\
    \multirow{2}[0]{*}{Serbia} & $-$0.001 & $-$0.001 & $-$0.001 \\
          & (0.05) & (0.05) & (0.05) \\
    \multirow{2}[0]{*}{$\overline{\mathit{AWP}}$ (Mean $\mathit{AWP}$)} & 1.338*** & 1.334*** & 1.350*** \\
          & (0.345) & (0.342) & (0.35) \\
    \multirow{2}[0]{*}{$\overline{\mathit{SoS}}$ (Mean $\mathit{SoS}$)} & 0.034 & 0.067 & 0.023 \\
          & (0.065) & (0.13) & (0.044) \\ \hline \showrowcolors
    Mundlak test & 16.138*** & 16.306*** & 15.963*** \\
    $R^2$ & 0.426 & 0.426 & 0.426 \\
    Adjusted $R^2$ & 0.417 & 0.417 & 0.417 \\
    AIC   & 453.624 & 453.583 & 453.679 \\
    BIC   & 481.494 & 481.453 & 481.548 \\
    RMSE  & 0.422 & 0.422 & 0.422 \\
    Observations & 396   & 396   & 396 \\ \toprule
    \end{tabularx}
\begin{tablenotes} \footnotesize
\item
\emph{Notes}:
The dependent variable is points per match.
Scotland is the reference category. Club-specific temporal means of $\mathit{AWP}$ and $\mathit{SoS}$ are included following a Mundlak-style within–between decomposition. Standard errors in parentheses are HC1 robust standard errors clustered at the club level. AIC = Akaike information criterion; BIC = Bayesian information criterion; RMSE = root mean square error.
Significance: * $p < 5\%$; ** $p < 1\%$; *** $p < 0.1\%$.
\end{tablenotes}
\end{threeparttable}
\end{table}

Table~\ref{Table7} presents the results of the baseline regression, which reveals a positive and statistically significant within-club association between $\mathit{AWP}$ and sporting performance. In particular, a 10 percentage point increase in $\mathit{AWP}$ for the same club is associated with an increase of approximately 0.16 points per match, \emph{ceteris paribus}.
In contrast, $\mathit{SoS}$ exhibits no statistically significant within-club association with $S$.
The coefficient of mean $\mathit{AWP}$ is also positive and statistically significant, whereas mean $\mathit{SoS}$ remains insignificant. The sum of the within-club $\mathit{AWP}$ coefficient and the mean-$\mathit{AWP}$ coefficient gives an implied between-club slope of approximately 2.935. Thus, both within-club variation and systematic differences across clubs contribute to the observed relationship between $\mathit{AWP}$ and sporting performance.
Club-specific temporal means are jointly significant, supporting their inclusion in the Mundlak-augmented pooled OLS specification.

Neither country indicator is statistically significant. Consequently, when performance is normalised and the scheduling variables are controlled for, there is no detectable conditional difference in points per match across Hungary, Scotland, and Serbia.
This is reinforced by the joint insignificance of $\mathit{AWP} \times \mathit{HUN}$ and $\mathit{AWP} \times \mathit{SRB}$ interactions ($\chi^2(2) = 0.012$, $p = 0.994$).

\subsection{Robustness checks} \label{Sec53}

Our baseline specification computes $\mathit{AWP}$ and $\mathit{SoS}$ with a home advantage of $\Delta = 50$, which might not hold as illustrated by Table~\ref{Table2}. To assess whether the results depend on this choice, the model is re-estimated with $\Delta = 25$ and $\Delta = 75$, see Table~\ref{Table7}. The estimated within-club AWP coefficients remain remarkably close to the baseline coefficient. The conclusions for $\mathit{SoS}$ are equally stable, its coefficient remains statistically insignificant in all specifications. The country indicators are also insignificant, while mean $\mathit{AWP}$ is positive and significant.
Therefore, the results are robust to the value of home advantage in terms of Elo ratings, with negligible changes in the coefficients.

Additional specification checks are conducted using the baseline model ($\Delta = 50$).
First, year indicators are included to account for common season-specific shocks. They are jointly insignificant ($\chi^2(10) = 0.848$, $p > 0.999$), and their inclusion essentially does not change the principal coefficients.
Second, because the three country panels do not have identical temporal coverage, the baseline model is re-estimated using only the nine common seasons (2015/16--2018/19 and 2021/22--2025/26), with club-specific temporal means recalculated for the restricted sample. The coefficient of $\mathit{AWP}$ remains significant ($\beta_1 = 1.723$, $p < 0.001$), while $\mathit{SoS}$ and the country indicators remain insignificant. Thus, our results are not driven by differences in temporal coverage.

\begin{table}[t!]
\centering
  \caption{Mundlak-augmented pooled OLS estimates, secondary sample (Scotland)}
  \label{Table8}
\begin{threeparttable}
    \rowcolors{1}{gray!20}{}
    \begin{tabularx}{\textwidth}{lCCC} \toprule \hiderowcolors
    Model & Baseline & Sensitivity & Sensitivity \\
    Home advantage $\Delta$ & 50    & 25    & 75 \\ \midrule
    \multirow{2}[0]{*}{Constant} & $-$0.221*** & $-$0.208*** & $-$0.243*** \\
          & (0.052) & (0.052) & (0.053) \\
    \multirow{2}[0]{*}{$\mathit{AWP}$} & 1.408*** & 1.394*** & 1.430*** \\
          & (0.276) & (0.273) & (0.28) \\
    \multirow{2}[0]{*}{$\mathit{SoS}$} & 0.048 & 0.097 & 0.032 \\
          & (0.039) & (0.077) & (0.026) \\
    \multirow{2}[0]{*}{$\overline{\mathit{AWP}}$ (Mean $\mathit{AWP}$)} & 1.790*** & 1.778*** & 1.812*** \\
          & (0.266) & (0.263) & (0.27) \\
    \multirow{2}[0]{*}{$\overline{\mathit{SoS}}$ (Mean $\mathit{SoS}$)} & 0.251 & 0.51  & 0.166 \\
          & (0.172) & (0.34) & (0.115) \\ \hline \showrowcolors
    Mundlak test & 51.852*** & 52.652*** & 51.236*** \\
    $R^2$ & 0.615 & 0.615 & 0.615 \\
    Adjusted $R^2$ & 0.609 & 0.610 & 0.609 \\
    AIC   & 248.016 & 247.775 & 248.110 \\
    BIC   & 266.535 & 266.294 & 266.629 \\
    RMSE  & 0.360 & 0.360 & 0.360 \\
    Observations & 300   & 300   & 300 \\ \toprule
    \end{tabularx}
\begin{tablenotes} \footnotesize
\item
\emph{Notes}:
The dependent variable is points per match.
Club-specific temporal means of $\mathit{AWP}$ and $\mathit{SoS}$ are included following a Mundlak-style within–between decomposition. Standard errors in parentheses are HC1 robust standard errors clustered at the club level. AIC = Akaike information criterion; BIC = Bayesian information criterion; RMSE = root mean square error.
Significance: * $p < 5\%$; ** $p < 1\%$; *** $p < 0.1\%$.
\end{tablenotes}
\end{threeparttable}
\end{table}

The analysis was repeated for the secondary sample, the Scottish Premiership from 2000/01 to 2025/26, excluding the 2019/20 season. The extended sample comprises 300 club-season observations, compared with 120 observations in the corresponding Scottish subsample of the primary sample. Both periods were estimated using the same Mundlak-augmented pooled OLS specification with club-clustered robust standard errors.

The results are consistent across the two observation periods. Both $\mathit{AWP}$ and mean $\mathit{AWP}$ remain positive and statistically significant with a small difference in their coefficients, whereas neither $\mathit{SoS}$ nor club mean $\mathit{SoS}$ is statistically significant. The joint Mundlak tests reject the null hypothesis that the coefficients of club-specific means are jointly zero ($p < 0.001$). The models explain approximately 61.5\% of the variation in points per match in contrast to about 42.6\% for the primary sample, indicating that the results of the Scottish Premiership are less uncertain than the Hungarian and Serbian leagues.

To summarise, excluding the pre-2015 Scottish observations from the primary cross-country analysis does not appear to alter the main findings. Even though the insignificant $\mathit{SoS}$ coefficient might be due to limited sample size, the impact of scheduling asymmetry remains below 0.1 points per match as the absolute value of $\mathit{SoS}$ never exceeds 1.5 (Table~\ref{Table3b}), which translates to less than one point (the value of a draw) in the single round-robin (sub)tournament containing 11 rounds.

\section{Discussion and conclusion} \label{Sec6}

A tournament with an even number of contestants may be unfair if the matches are asymmetric due to some reasons, such as playing with the white pieces in chess or playing at home in most team sports.
Therefore, our paper has empirically investigated the fairness of single round-robin tournaments. To that end, we have constructed a unique dataset from the first division of three European football leagues, where a standard double round-robin tournament is continued with a single round-robin (sub)tournament. This setting allows an accurate identification of team strengths, which is crucial to analyse whether the allocation of home games provides a detectable (dis)advantage in the number of points collected. We have also developed a strength of schedule metric using Elo ratings that has favourable theoretical properties.

The research question has been studied with a Mundlak-augmented pooled OLS regression that bridges standard random-effects and fixed-effects panel models. The statistical estimations show a consistent picture: while the average winning probability has a clear and robust positive effect on the number of points per match, the strength of schedule does not impact the points at all. In the primary sample, the coefficient is even negative, suggesting that the insignificant result is not caused by limited sample size. In the secondary sample of the Scottish league, the strength of schedule has a positive but insignificant effect, with a negligible (if any) impact on the final ranking.

This finding may be surprising at first sight. After all, home advantage is well-documented in football, and can also be seen in the descriptive statistics of the sample. \citet{CsatoKrumer2026} have recently examined a similar problem: the fairness of Swiss-system chess tournaments containing an odd number of rounds. There, players with more white games enjoy a substantial, statistically significant advantage compared to their peers with more black games. Due to the analogy between home advantage in team sports and white advantage in chess \citep{vanOurs2024}, one could anticipate a positive impact in our sample, too -- but the results presented above do not support the straightforward expectation. Consequently, using the rather complex ranking-fair schedules proposed by \citet{WesselHurkensSpieksma2026} seems to be unnecessary in the football leagues considered here.

However, the lack of empirical evidence does not guarantee the fairness of these tournaments. A possible reason is that home advantage is quantified by adding the same constant to the Elo ratings of the home team in each match. Even though all widely used Elo variants, such as FIFA Women's World Ranking \citep{vanEetveldeLey2019}, Football Club Elo ratings, and World Football Elo Ratings (\url{https://eloratings.net/}) \citep{GasquezRoyuela2016, Csato2022a}, account for home advantage in this way, home advantage has a team-specific component \citep{PeetersvanOurs2021}, which cannot be captured by this approach. Non-transitive patterns in long-term rivalries \citep{vanOurs2024}, bogey team relationships \citep{vanOurs2025b}, and circular triads of overperformance \citep{vanOurs2026} could also interact with the allocation of the field of the games, and may lead to unfairness when some teams benefit from a given set of home opponents.
Hopefully, our study will inspire future research on single round-robin tournaments from these perspectives.

\section*{Acknowledgements}
\addcontentsline{toc}{section}{Acknowledgements}

%We are grateful to \emph{David Winkelmann} for useful remarks. \\
%and three anonymous reviewers
\noindent
The research was supported by the National Research, Development and Innovation Office under Grants Advanced 152220 and FK 145838, and the J\'anos Bolyai Research Scholarship of the Hungarian Academy of Sciences.

\bibliographystyle{apalike} 
\bibliography{All_references}

\clearpage
\setcounter{section}{0}
\renewcommand{\thesubsection}{\Alph{subsection}}

\setcounter{figure}{0}
\renewcommand{\thefigure}{A.\arabic{figure}}

\setcounter{table}{0}
\renewcommand{\thetable}{A.\arabic{table}}

\section*{Appendix}
\addcontentsline{toc}{section}{Appendix}

\begin{table}[ht!]
  \centering
  \caption{First matchdays of the single round-robin (sub)tournaments}
  \label{Table_A1}

\begin{threeparttable}
    \rowcolors{1}{gray!20}{}
    \begin{tabularx}{0.6\textwidth}{lCCC} \toprule \hiderowcolors
    \multirow{2}[0]{*}{Season} & \multicolumn{3}{c}{Country} \\
          & Hungary & Scotland & Serbia \\ \bottomrule \showrowcolors
    2000/01 & ---   & 2000-12-26 & --- \\
    2001/02 & ---   & 2002-01-12 & --- \\
    2002/03 & ---   & 2002-12-29 & --- \\
    2003/04 & ---   & 2004-02-10 & --- \\
    2004/05 & ---   & 2005-01-22 & --- \\
    2005/06 & ---   & 2006-01-21 & --- \\
    2006/07 & ---   & 2007-01-01 & --- \\
    2007/08 & ---   & 2008-01-19 & --- \\
    2008/09 & ---   & 2009-01-24 & --- \\
    2009/10 & ---   & 2010-01-30 & --- \\
    2010/11 & ---   & 2011-01-22 & --- \\
    2011/12 & ---   & 2012-01-14 & --- \\
    2012/13 & ---   & 2013-01-19 & --- \\
    2013/14 & ---   & 2014-01-18 & --- \\
    2014/15 & ---   & 2015-01-20 & --- \\
    2015/16 & 2016-03-05 & 2016-01-15 & 2016-04-15 \\
    2016/17 & 2017-03-11 & 2017-01-31 & 2017-04-14 \\
    2017/18 & 2018-03-17 & 2018-01-23 & 2018-04-13 \\
    2018/19 & 2019-03-02 & 2019-01-26 & 2019-04-12 \\
    2019/20 & 2020-02-29 & ---   & --- \\
    2020/21 & 2021-02-26 & 2021-01-09 & --- \\
    2021/22 & 2022-03-04 & 2022-01-25 & 2022-04-15 \\
    2022/23 & 2023-03-11 & 2023-01-28 & 2023-04-21 \\
    2023/24 & 2024-03-02 & 2024-01-27 & 2024-04-20 \\
    2024/25 & 2025-03-07 & 2025-01-11 & 2025-04-12 \\
    2025/26 & 2026-02-21 & 2026-01-24 & 2026-04-18 \\ \bottomrule
    \end{tabularx}
\begin{tablenotes} \footnotesize
\item
\emph{Note}:
Football Club Elo Ratings can be downloaded from \url{api.clubelo.com/YEAR-MM-DD}.
\end{tablenotes}
\end{threeparttable}
\end{table}

Table~\ref{Table_A1} presents the first day of Round 23 in the Hungarian and Scottish leagues, as well as the first day of Round 31 in the Serbian league. Elo ratings on these days are used in the statistical analysis.

\subsection{Data corrections and remarks}

Some adjustments have been necessary in the Elo ratings and match results.

In Hungary, the league uses a mirrored schedule \citep{GoossensSpieksma2012b}:  the second half of the double round-robin subtournament is identical to the first, except that the venues are inverted, and the third part copies the first half of the double round-robin subtournament.
However, in the 2019/20 season, the two matches Kaposv\'ar vs \'Ujpest took place in Rounds 11 and 22, while \'Ujpest vs Kaposv\'ar took place in Round 33. Hence, the result of Kaposv\'ar vs \'Ujpest in Round 22 (0-1) was used to calculate the number of points collected in Rounds 23--33 instead of the result of \'Ujpest vs Kaposv\'ar, and the Elo ratings of 22 February 2020 were used for Kaposv\'ar and \'Ujpest regarding this particular match in the calculation of average winning probability and strength of schedule.
Analogously, in the 2020/21 season, the two matches Feh\'erv\'ar vs Pusk\'as Akad\'emia took place in Rounds 4 and 15, while Pusk\'as Akad\'emia vs Feh\'erv\'ar took place in Round 26. Hence, the result of Feh\'erv\'ar vs Pusk\'as Akad\'emia in Round 15 (1-1) was used to calculate the number of points collected in Rounds 23--33 instead of the result of Pusk\'as Akad\'emia vs Feh\'erv\'ar, and the Elo ratings of 15 December 2020 were used for Feh\'erv\'ar and Pusk\'as Akad\'emia regarding this particular match in the calculation of average winning probability and strength of schedule.

In the 2023/24 Scottish Premiership, the match Dundee vs Kilmarnock in Round 23 was played on 3 November 2024, even though Round 23 started only on 11 January 2025 as shown in Table~\ref{Table_A1}.  Consequently, the Elo ratings of 3 November 2024 were used for Dundee and Kilmarnock regarding this particular match in the calculation of average winning probability and strength of schedule.

The Serbian club Javor Ivanjica/Javor Matis played in the first division in five out of the nine seasons in our sample. Between 2015/16 and 2017/18, it is called Loznica in Football Club Elo Ratings; between 2022/23 and 2023/24, it is called Ivanjica in Football Club Elo Ratings; thus, Football Club Elo Ratings consider it two separate teams.

\end{document}

%% file: Figure1_AWP_points_per_match.tex
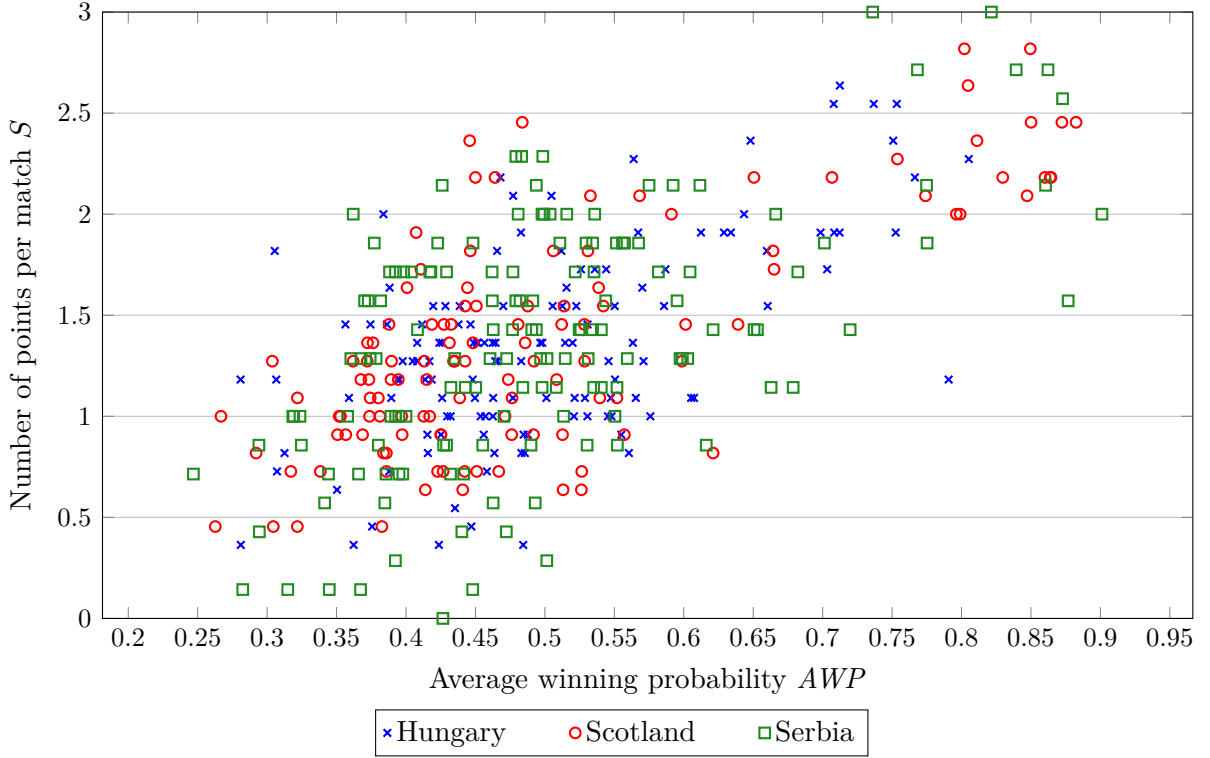
\begin{figure}[t!]
\centering

\begin{tikzpicture}
\begin{axis}[
name = axis1,
xlabel = Average winning probability $\mathit{AWP}$,
x label style = {font=\small},
x tick label style = {/pgf/number format/1000 sep={}},
ylabel = Number of points per match $S$,
y label style = {font=\small},
width = \textwidth,
height = 0.6\textwidth,
ymajorgrids = true,
ymin = 0,
ymax = 3,
%max space between ticks=50,
legend style = {font=\small,at={(0.25,-0.15)},anchor=north west,legend columns=3},
legend entries = {Hungary$\qquad$,Scotland$\qquad$,Serbia}
]
% Hungary, 2015-2026
\addplot [blue, mark=x, mark size=2pt, mark options=solid, only marks, thick] coordinates {
(0.280878161753371,1.18181818181818)
(0.628944704643416,1.90909090909091)
(0.464487791909293,1.36363636363636)
(0.752631595347732,1.90909090909091)
(0.448688269265008,1.36363636363636)
(0.446416847320435,1.45454545454545)
(0.54751614444075,1.09090909090909)
(0.555042230695708,0.909090909090909)
(0.358923803414328,1.09090909090909)
(0.575946774729454,1)
(0.306566960986415,1.18181818181818)
(0.633956715494091,1.90909090909091)
(0.505590743868702,1.54545454545455)
(0.386326386744368,1.45454545454545)
(0.660466345645592,1.54545454545455)
(0.312527771350452,0.818181818181818)
(0.563928278733446,2.27272727272727)
(0.42517777184589,0.909090909090909)
(0.465947739081949,1.27272727272727)
(0.514618776050666,1.36363636363636)
(0.462555262592053,1)
(0.521421614236829,1.09090909090909)
(0.482941522083326,1.27272727272727)
(0.698497787766727,1.90909090909091)
(0.393596600941156,1.18181818181818)
(0.52887163175375,1.09090909090909)
(0.458227141331223,0.727272727272727)
(0.703249913211397,1.72727272727273)
(0.408052519136423,1.36363636363636)
(0.544116023466923,1.72727272727273)
(0.404892693057679,1.27272727272727)
(0.530633549644842,1)
(0.388193104419427,1.63636363636364)
(0.512547189819557,1.54545454545455)
(0.415399396807621,0.909090909090909)
(0.712220236410001,1.90909090909091)
(0.522625441200577,1.54545454545455)
(0.408108428432881,1.27272727272727)
(0.707946642952848,2.54545454545455)
(0.305470169192245,1.81818181818182)
(0.545866179187259,1.27272727272727)
(0.397611752353631,1.27272727272727)
(0.451675504205065,1.36363636363636)
(0.423684214301734,0.363636363636364)
(0.485580700240784,0.818181818181818)
(0.43778713294059,1.45454545454545)
(0.60532110326075,1.09090909090909)
(0.708322731731637,1.90909090909091)
(0.547958844788909,1)
(0.477004138161644,1.09090909090909)
(0.457138757188105,1)
(0.659876644402198,1.81818181818182)
(0.766428585667634,2.18181818181818)
(0.281048013106564,0.363636363636364)
(0.41960616897103,1.54545454545455)
(0.560475635774464,0.818181818181818)
(0.424034305029367,1.36363636363636)
(0.525973786793066,1.72727272727273)
(0.49671515561157,1.36363636363636)
(0.383739964505449,2)
(0.362264392225475,0.363636363636364)
(0.374402167281841,1.45454545454545)
(0.612398412706607,1.90909090909091)
(0.805228693836327,2.27272727272727)
(0.462478503341817,1.36363636363636)
(0.451219200471034,1.36363636363636)
(0.501060519768966,1.09090909090909)
(0.463649839959596,0.818181818181818)
(0.515580360787641,1.63636363636364)
(0.550262913132382,1.54545454545455)
(0.438683682939872,1.54545454545455)
(0.462771313548441,1.09090909090909)
(0.429706474585137,1)
(0.535948770238861,1.72727272727273)
(0.753412416812304,2.54545454545455)
(0.387923820193875,0.727272727272727)
(0.448046609232336,1.18181818181818)
(0.570132170593882,1.63636363636364)
(0.453792218886575,1)
(0.356366391958031,1.45454545454545)
(0.565464788200165,1.09090909090909)
(0.550524934606047,1.18181818181818)
(0.465605107401421,1.81818181818182)
(0.483076297291367,0.818181818181818)
(0.477151030163187,2.09090909090909)
(0.520902068822728,1)
(0.790750551718806,1.18181818181818)
(0.432134576110175,1)
(0.482786957409324,1.90909090909091)
(0.545406025467172,1)
(0.464262185040253,1.27272727272727)
(0.511865409212813,1.81818181818182)
(0.504747441531541,2.09090909090909)
(0.470036885120939,1.54545454545455)
(0.350322276764994,0.636363636363636)
(0.449634592638067,1.09090909090909)
(0.750834995571956,2.36363636363636)
(0.607415017317217,1.09090909090909)
(0.563379253183111,1.36363636363636)
(0.567162187476872,1.90909090909091)
(0.530711835372019,1.45454545454545)
(0.418506393642475,1.18181818181818)
(0.498099643090147,1.36363636363636)
(0.456492930193572,1.36363636363636)
(0.428385828211035,1.54545454545455)
(0.395322707288764,1.18181818181818)
(0.408099636563014,1.27272727272727)
(0.375589572089819,0.454545454545455)
(0.426009859945507,1.36363636363636)
(0.484684496179706,0.909090909090909)
(0.484386379157259,0.363636363636364)
(0.712387220132061,2.63636363636364)
(0.468005471487157,2.18181818181818)
(0.435272788884762,0.545454545454545)
(0.487666015011164,0.909090909090909)
(0.389404083308509,1.09090909090909)
(0.586787192622957,1.72727272727273)
(0.643413408469574,2)
(0.426009859945507,1.09090909090909)
(0.455973224855837,0.909090909090909)
(0.519942624492973,1.36363636363636)
(0.447054295954828,0.454545454545455)
(0.736807280698468,2.54545454545455)
(0.648058463958889,2.36363636363636)
(0.307105529447889,0.727272727272727)
(0.41573983536008,0.818181818181818)
(0.413735100011054,1.18181818181818)
(0.411589593898386,1.45454545454545)
(0.585809629654623,1.54545454545455)
(0.571048315869925,1.27272727272727)
(0.416886359435617,1.27272727272727)
};
% Scotland, 2015-2026
\addplot [red, mark=o, mark size=2pt, mark options=solid, only marks, thick] coordinates {
(0.81113237283862,2.36363636363636)
(0.435225849084891,1.27272727272727)
(0.372077759406452,1.27272727272727)
(0.38921303292065,1.18181818181818)
(0.506060796690041,1.81818181818182)
(0.539559591284612,1.09090909090909)
(0.351907735402257,1)
(0.432500063242481,1.45454545454545)
(0.471288738850572,1)
(0.47360534285883,1.18181818181818)
(0.551900235060445,1.09090909090909)
(0.650554794239749,2.18181818181818)
(0.8722833420284,2.45454545454545)
(0.422745854023326,0.727272727272727)
(0.37225196031824,1.36363636363636)
(0.528631071718996,1.27272727272727)
(0.44222670611519,0.727272727272727)
(0.374180305418588,1.09090909090909)
(0.417083967162622,1)
(0.444223912679201,1.63636363636364)
(0.512084998967621,1.45454545454545)
(0.425014237191844,0.909090909090909)
(0.538718850136223,1.63636363636364)
(0.66501483211091,1.72727272727273)
(0.863839939433956,2.18181818181818)
(0.38585389104166,0.727272727272727)
(0.413906199711721,0.636363636363636)
(0.528376162236176,1.45454545454545)
(0.464103574407057,2.18181818181818)
(0.445874831212158,2.36363636363636)
(0.396965914548894,1)
(0.382673747232747,0.454545454545455)
(0.591161718125399,2)
(0.38585389104166,0.818181818181818)
(0.476375298897663,1.09090909090909)
(0.639100928100806,1.45454545454545)
(0.804695218225038,2.63636363636364)
(0.304438138083699,0.454545454545455)
(0.266784292137715,1)
(0.492159628272082,1.27272727272727)
(0.532791004683984,2.09090909090909)
(0.598734183698341,1.27272727272727)
(0.438691827054472,1.09090909090909)
(0.410713260239579,1.72727272727273)
(0.706725007553727,2.18181818181818)
(0.513116991083384,0.636363636363636)
(0.292049520867172,0.818181818181818)
(0.621145556887467,0.818181818181818)
(0.796249302779908,2)
(0.367476432490518,1.18181818181818)
(0.33830667623276,0.727272727272727)
(0.53095871014518,1.81818181818182)
(0.44083160591182,0.636363636363636)
(0.485819689517829,1.36363636363636)
(0.413185373881519,1.27272727272727)
(0.882494745178457,2.45454545454545)
(0.321796794744233,1.09090909090909)
(0.40728221366693,1.90909090909091)
(0.394452898563378,1.18181818181818)
(0.526282941828751,0.636363636363636)
(0.802073405627403,2.81818181818182)
(0.31707408811352,0.727272727272727)
(0.376258288251257,1.36363636363636)
(0.487685252465664,1.54545454545455)
(0.526557207233607,0.727272727272727)
(0.448245176703281,1.36363636363636)
(0.466783260844126,0.727272727272727)
(0.860070515755117,2.18181818181818)
(0.400729957273339,1.63636363636364)
(0.373308788966922,1.18181818181818)
(0.414931116937013,1.18181818181818)
(0.449987784668535,2.18181818181818)
(0.849559460437542,2.81818181818182)
(0.425220552035168,0.909090909090909)
(0.557222501171933,0.909090909090909)
(0.450673309721022,1.54545454545455)
(0.352844082227558,1)
(0.491808545156805,0.909090909090909)
(0.38955482443273,1.27272727272727)
(0.850106407710122,2.45454545454545)
(0.356656597007838,0.909090909090909)
(0.383778385817307,0.818181818181818)
(0.442587549613439,1.54545454545455)
(0.512693504512601,0.909090909090909)
(0.829691946777326,2.18181818181818)
(0.380094961036313,1.09090909090909)
(0.568205317601914,2.09090909090909)
(0.50846817186716,1.18181818181818)
(0.446329726931681,1.81818181818182)
(0.321759668526139,0.454545454545455)
(0.418643939357787,1.45454545454545)
(0.847210572588392,2.09090909090909)
(0.350677131216552,0.909090909090909)
(0.373724327163872,1)
(0.442500732420264,1.27272727272727)
(0.480784946942909,1.45454545454545)
(0.864409816222098,2.18181818181818)
(0.397238960670943,0.909090909090909)
(0.42727715298851,1.45454545454545)
(0.513898017315413,1.54545454545455)
(0.483700439198611,2.45454545454545)
(0.476107102946354,0.909090909090909)
(0.450936618052368,0.727272727272727)
(0.799000445209231,2)
(0.368718959933228,0.909090909090909)
(0.303745810511941,1.27272727272727)
(0.434181730008395,1.27272727272727)
(0.426749394073359,0.727272727272727)
(0.774056721762786,2.09090909090909)
(0.381196873245064,1)
(0.431277693200397,1.36363636363636)
(0.387765509958046,1.45454545454545)
(0.664309399606741,1.81818181818182)
(0.601324898115895,1.45454545454545)
(0.361941895345919,1.27272727272727)
(0.26264734586209,0.454545454545455)
(0.542092208513694,1.54545454545455)
(0.753832119492416,2.27272727272727)
(0.412805940823593,1)
};
% Serbia, 2015-2026
\addplot [ForestGreen, mark=square, mark size=2pt, mark options=solid, only marks, thick] coordinates {
(0.389156706969889,1)
(0.521829244648872,1.71428571428571)
(0.701275445166267,1.85714285714286)
(0.392418224710236,1.71428571428571)
(0.324645295011462,0.857142857142857)
(0.775226770752522,1.85714285714286)
(0.529629263627673,1.85714285714286)
(0.365819049113078,0.714285714285714)
(0.501365994720971,0.285714285714286)
(0.498003207358981,1.14285714285714)
(0.525901901794352,1.42857142857143)
(0.497938848387655,2)
(0.53116651032361,1.28571428571429)
(0.462753562640568,0.571428571428571)
(0.50376334562457,2)
(0.479106629149293,2.28571428571429)
(0.385591558248403,0.714285714285714)
(0.567472730681108,1.85714285714286)
(0.344754419680341,0.142857142857143)
(0.398254942855981,1.71428571428571)
(0.384765684891279,0.571428571428571)
(0.736052384192081,3)
(0.408274496388801,1.42857142857143)
(0.774833783062005,2.14285714285714)
(0.435076283105148,1.28571428571429)
(0.448242320180504,1.85714285714286)
(0.470565737694682,1)
(0.400008206663705,1)
(0.550449088116197,1)
(0.461977478104604,1.71428571428571)
(0.551526719823988,1.85714285714286)
(0.682154166311172,1.71428571428571)
(0.455267189294614,0.857142857142857)
(0.392538506463117,1)
(0.719791560557329,1.42857142857143)
(0.417176894803914,1.71428571428571)
(0.821504277665552,3)
(0.388137306327631,1.71428571428571)
(0.427086473884559,0.857142857142857)
(0.378497791003284,1.28571428571429)
(0.462826669296603,1.42857142857143)
(0.440232310206314,0.428571428571429)
(0.491292423962596,1.57142857142857)
(0.490558570890211,1.42857142857143)
(0.602858063821419,1.28571428571429)
(0.426045566256799,2.14285714285714)
(0.534271255967383,1.42857142857143)
(0.551915139598674,1.14285714285714)
(0.476998008797367,1.71428571428571)
(0.372812448035094,1.57142857142857)
(0.357946694815885,1)
(0.611634765614258,2.14285714285714)
(0.294385329461862,0.428571428571429)
(0.62110819389457,1.42857142857143)
(0.872768210306746,2.57142857142857)
(0.392346349074217,0.285714285714286)
(0.432051690720481,0.714285714285714)
(0.377153578640367,1.85714285714286)
(0.514639388344079,1.28571428571429)
(0.417927360587981,1.71428571428571)
(0.559338459672479,1.28571428571429)
(0.67877277651589,1.14285714285714)
(0.616188674997167,0.857142857142857)
(0.403928070521557,1.71428571428571)
(0.862254751364773,2.71428571428571)
(0.768264440272035,2.71428571428571)
(0.53039936433044,0.857142857142857)
(0.381747771211356,1.57142857142857)
(0.367569483172698,1.28571428571429)
(0.394816147647133,1)
(0.380100852752426,0.857142857142857)
(0.314847189249139,0.142857142857143)
(0.592304601413537,2.14285714285714)
(0.429207664361024,1.71428571428571)
(0.597344068067674,1.28571428571429)
(0.540663973092189,1.42857142857143)
(0.48330496935392,2.28571428571429)
(0.394967143164873,0.714285714285714)
(0.490003752742852,0.857142857142857)
(0.472203827803931,0.428571428571429)
(0.86051804612987,2.14285714285714)
(0.510789523584486,1.85714285714286)
(0.557339105606954,1.85714285714286)
(0.666218565116945,2)
(0.472469836550361,1.28571428571429)
(0.367274079163533,0.142857142857143)
(0.318505037194661,1)
(0.246885806653189,0.714285714285714)
(0.508254963987519,1.14285714285714)
(0.480761774023426,2)
(0.493830662175979,2.14285714285714)
(0.442711810992102,1.14285714285714)
(0.493027586983944,0.571428571428571)
(0.552191682934254,0.857142857142857)
(0.535390856726797,1.71428571428571)
(0.493830662175979,1.42857142857143)
(0.839350656267506,2.71428571428571)
(0.662907027912828,1.14285714285714)
(0.575006255696103,2.14285714285714)
(0.479085798423157,1.57142857142857)
(0.370181532591728,1.57142857142857)
(0.496949363868031,1.28571428571429)
(0.294064061799598,0.857142857142857)
(0.28245530344105,0.142857142857143)
(0.581741270247434,1.71428571428571)
(0.53435987917374,1.85714285714286)
(0.476743922967965,1.42857142857143)
(0.540629259674248,1.14285714285714)
(0.46052579844947,1.28571428571429)
(0.422830933502396,1.85714285714286)
(0.535140043381186,1.14285714285714)
(0.448028892603561,0.142857142857143)
(0.901133507730683,2)
(0.653060598789253,1.42857142857143)
(0.361995258588581,2)
(0.323650501784414,1)
(0.432308148851436,1.14285714285714)
(0.482119372694052,1.57142857142857)
(0.501552526093172,1.28571428571429)
(0.344180085468409,0.714285714285714)
(0.650622801967685,1.42857142857143)
(0.51577561179902,2)
(0.543828080036194,1.57142857142857)
(0.524684146498047,1.42857142857143)
(0.462172361021313,1.57142857142857)
(0.51345642407318,1)
(0.429190440886473,0.857142857142857)
(0.360270133718088,1.28571428571429)
(0.876804128422163,1.57142857142857)
(0.535878655892371,2)
(0.604531511017408,1.71428571428571)
(0.450189364669298,1.14285714285714)
(0.441529672883318,0.714285714285714)
(0.374193696601992,1.28571428571429)
(0.319151455917122,1)
(0.397721514596328,0.714285714285714)
(0.555886322516558,1.85714285714286)
(0.498503055548542,2.28571428571429)
(0.499323565089226,2)
(0.599005465656147,1.28571428571429)
(0.595198935184134,1.57142857142857)
(0.48411707985644,1.14285714285714)
(0.426563379078788,0)
(0.341402197070164,0.571428571428571)
};
% Linear trend extra black
%\draw [black,thick,dotted] (\pgfkeysvalueof{/pgfplots/xmin},-0.6008+31.522*\pgfkeysvalueof{/pgfplots/xmin}) -- (\pgfkeysvalueof{/pgfplots/xmax},-0.6008+31.522*\pgfkeysvalueof{/pgfplots/xmax});
\end{axis}
\end{tikzpicture}

\captionsetup{justification=centering}
\caption{The connection between average winning probability and \\ the number of points per match, primary sample (2015/16--2025/26)}
\label{Fig1}
\end{figure}

%\end{document}